\documentstyle[epsf,12pt]{article}
\begin{document}

\def\lpmb#1{\mbox{\boldmath$#1$}}

\centerline{ \bf Computational Advances in 
  the Study of Baryon Number Violation}
\centerline{\bf in High Energy Electroweak Collisions
\footnote[1]{This paper  is based on a talk given by R.
  Singleton at the 9th International Seminar ``Quarks-96'' 
  held in Yaroslavl, Russia, and will be published the 
  proceedings of this conference.}
}

\vskip0.5cm
\centerline{Claudio Rebbi\footnote[2]{rebbi@pthind.bu.edu} 
and Robert Singleton, Jr.\footnote[3]{bobs@cthulu.bu.edu}}
\smallskip
\centerline{\it Physics Department}
\centerline{\it Boston University}
\centerline{\it 590 Commonwealth Avenue}
\centerline{\it Boston, MA 02215, USA}

\setcounter{page}{0}
\thispagestyle{empty}

\vfill

\centerline{\bf Abstract}
\vskip0.2in

We present some recent advances in the computational
study of baryon number violation in high energy
electroweak collisions. We examine classically allowed 
processes above the sphaleron barrier and, using a 
stochastic search procedure, we explore the topology 
changing region in the energy and particle number plane. 
Finding topology changing classical solutions with small 
incident particle number would be an indication that
baryon number violation becomes unsuppressed in high 
energy collisions. Starting with a topology changing 
solution of approximately 50 incoming particles, our 
Monte-Carlo procedure has produced other topology changing 
solutions with 40\% lower incident particle numbers, 
with energies up to one and a half times the sphaleron 
energy. While these solutions still involve a rather 
large number of incident particles, we have nonetheless
demonstrated that our search procedure is effective 
in reducing the particle number while ensuring
topology change. Taking advantage of more powerful
computational resources, we plan to extend the search
to still higher energies.
\vfill

\noindent BUHEP-96-13 

\noindent hep-ph/9606479 \hfill Typeset in La\TeX
\eject

\section{Introduction}
\hspace{1cm}

The prospect of observable high energy baryon number 
violation within the standard model has recently 
attracted widespread attention. Unfortunately, despite 
considerable effort by a great many theorists, the 
issue still remains largely unsettled. The purpose of 
this paper is to explain some recent computational
developments that shed more light on the problem 
and which might help contribute to a final solution.
Before we begin, however, in an effort to write a
self-contained work, we shall give a brief 
exposition of nonperturbative baryon number 
violation in the standard model. We unfortunately 
cannot survey  the vast literature on the subject 
with the depth it deserves, so instead a brief 
summary of the facts germane to our 
numerical approach must suffice.

For our purposes, when we talk of the
``standard model'' we mean the standard model 
in which the Weinberg angle has been set to zero,
i.e. we shall be considering $SU(2)$ gauge
theory spontaneously broken via a single Higgs 
doublet. This simplified model has all the
relevant physics. Most importantly, the 
gauge structure dictates nontrivial topology 
for the bosonic vacuum sector. Working in
the temporal gauge with periodic boundary
conditions at spatial infinity, each
vacuum may be characterized by
an integer called the winding number which
measures the number of times the gauge
manifold is wound around 3-space\cite{JR}. 
As this number is a topological invariant, 
vacua of different winding numbers 
cannot be continuously deformed into 
one another. 

Because of the axial vector anomaly, baryon number
violation occurs when the gauge and Higgs fields
change their topology\cite{thooft76}. Different 
topological sectors are separated by an extremely 
high barrier, the top of which is a static saddle-point 
solution to the equations of motion. This configuration 
is called the sphaleron\cite{KM}, and it has an energy
of about $10 \, {\rm TeV}$ and possesses a single 
unstable direction in field space. At low energy the 
baryon number violating rates are exceedingly small, 
as the gauge and Higgs fields must  first tunnel 
through the sphaleron, which is extremely unlikely 
indeed. 

Recently the prospect of rapid baryon number
violation at high temperatures and high energies 
has emerged. The basic idea is that if the gauge
and Higgs fields have enough energy, they can
change their topology by sailing over the sphaleron 
barrier rather than being forced to tunnel through 
it. At high temperatures this is precisely what
happens, and it is generally agreed that baryon
number violation becomes unsuppressed in the
early universe\cite{highT}. 

The situation in high energy collisions is far 
less clear. The limiting process in baryon 
number violation is the production of a 
sphaleron-like configuration from an
incident beam of high energy particles. 
But since the sphaleron 
is a large extended object, there is a scale 
mismatch with the initial high energy 
\hbox{2-particle} state, and hence one naively
expects the baryon number violating rate to be 
small. However, Ringwald
\cite{ring} and Espinosa\cite{esp} have suggested
that the sum over many-particle final states gives
rise to factors that grow with energy sufficiently
rapidly to offset any exponential suppression.
If true, this offers the exciting prospect of 
one day studying baryon number violation in the 
laboratory. Their approach, however, neglects some 
important corrections which still elude calculation 
despite considerable effort. Apart from lattice 
simulations, semi-classical techniques are our only 
handle on nonperturbative effects. The basic problem 
with anomalous baryon number violation is that exclusive
two-particle initial states are not very amenable
to these methods, and there are potentially large
initial state corrections whose effects
remained undetermined.

Our efforts lie in an attempt to overcome these
limitations. Many people have struggled in similar 
endeavors, but we only have space to summarize the 
work of one group upon which our approach 
has been partly inspired. In an effort to alleviate 
problems arising from exclusive two-particle 
states, Rubakov, Son and Tinyakov\cite{rst} 
consider an inclusive quantity: 
\begin{eqnarray}
\label{sen}
  \sigma(E,N) &=&   \sum_{f,i} \, 
  \mid<f| \, S \, P_E\, P_N \, |i>\mid^2 \, 
\end{eqnarray}
where the sum is over all initial and final states, $S$ is
the $S$-matrix, and $P_E$ and $P_N$ are projection
operators onto subspaces of energy $E$ and particle
number $N$ respectively.  

Unlike the exclusive two-particle 
cross section $\sigma_2(E)$, the quantity  $\sigma(E,N)$ 
is directly calculable by semiclassical methods. If the 
energy and particle number are
parameterized by
\begin{eqnarray}
\label{Egsq} 
  E &=& {\epsilon \over g^2} 
\\
\label{Ngsq}
  N &=& {\nu \over g^2} \ ,
\end{eqnarray}
then in the limit $g \to 0$ with $\epsilon$ and $\nu$
held fixed, the path integral for $\sigma(E,N)$ can 
be saturated by a classical saddle-point solution 
to the equations of motion. The cross section takes 
the form
\begin{eqnarray}
  \sigma(E,N) &=&  
  \exp \left[ {1 \over g^2} \, 
  F(\epsilon,\nu) + {\cal O}(g^0)\right]  \ ,
\end{eqnarray}
where the function $F(\epsilon,\nu)$ is determined
by the classical solution. These solutions naturally
divide into two regimes: there is Euclidean 
evolution corresponding to tunneling under the 
sphaleron barrier, and  Minkowski evolution 
corresponding to classical motion before and after 
the tunneling event. 

The utility of $\sigma(E,N)$ arises because it may be 
used to bound $\sigma_2(E)$. By construction, $\sigma(E,N)$
provides an upper bound to $\sigma_2(E)$. A lower
bound may be obtained under some reasonable 
physical assumptions. The two-particle process is
expected to be dominated by a most favored intermediate
sphaleron-like state, and the rate into this intermediate 
state bounds the two-particle  cross section from below.
Combining these upper and lower bounds 
allows one to write \cite{{pt},{modproj}}
\begin{eqnarray}
\label{sigineq}
  \exp(-const \, N) \, \sigma(E,N) < \sigma_2(E) \,
  < \sigma(E,N) \ ,
\end{eqnarray}
from which it follows that\cite{pt}
\begin{eqnarray}
  \lim_{g \to 0} \sigma_2(E) = F(\epsilon,\nu) + \cal{O(\nu)} \ .
\end{eqnarray}
The consistency of the first inequality requires
that $F(\epsilon,\nu)$ have a smooth $\nu \to 0$ 
limit, in which case $F(\epsilon,0)$ determines 
the exponential behavior of $\sigma_2(E)$. 
However,  the second inequality of (\ref{sigineq}) 
holds regardless of continuity, and hence
if $\sigma(E,N)$ is exponentially suppressed then
so is $\sigma_2(E)$.

Obviously finding these Euclidean-Minkowski hybrid 
solutions will be extremely illuminating, 
and we are presently engaged in this rather 
formidable numerical task. In this paper, 
however, rather than exploring the full 
barrier penetration problem, we report on 
a complimentary approach. We examine the 
classically allowed regime above the sphaleron 
barrier in which the saddle-points that 
saturate the path integral are pure Minkowski 
solutions. This is less computationally 
demanding than solving the tunneling problem,
while still yielding considerable information 
about baryon number violation. 
Spatial limitations prevent us from giving 
a full blown treatment of our numerical 
investigation, and the reader is referred 
to Ref.~\cite{rs} for complete details. 
But the basic idea is that if a topology 
changing classical solution with small incident 
particle number could be found, this would be 
a strong indication that baryon number violation 
would be observable in high energy two-particle 
collisions. Conversely, if there are no 
small-multiplicity topology changing solutions, 
then it is unlikely that the rates become 
exponentially unsuppressed.  

This can be made more precise in the following 
manner. Because of energy dissipation, the 
system will asymptotically approach vacuum 
values and will consequently linearize in 
the past and future. Field evolution then
becomes a superposition of normal mode 
oscillators with amplitudes $a_n$, which 
allows us to define the asymptotic particle 
number $\nu = \sum |a_n|^2$. 
Furthermore, since the fields approach 
vacuum values in the infinite past and 
future, the winding numbers of the asymptotic 
field configurations are also well defined, 
and fermion number violation is given by 
the change in topology of these vacua\cite{fggrs}. 
Because of the sphaleron barrier, classical
solutions that change topology must have
energy $\epsilon$ greater than that of the
sphaleron. The problem we would like to solve, 
then, is whether the incident particle number 
$\nu$ of these solutions can be made arbitrarily
small. That is to say, we wish to map the
region of topology changing classical solutions
in the $\epsilon$-$\nu$ plane. 

We could easily parameterize incoming
configurations in terms of small perturbations 
about a given vacuum, but it would be 
extremely difficult to choose the parameters 
to ensure a subsequent change in winding
number. This is because topology changing 
classical solutions must pass over the sphaleron 
barrier at some point in their evolution, which 
is extremely difficult arrange by an 
appropriate choice of initial conditions. 
So computationally
we purse a different strategy. We shall evolve
a configuration near the top of the sphaleron
barrier until it linearizes and the particle number
can be extracted. The time reversed solution,
then, has a known incident particle number and 
will typically pass over the sphaleron barrier
thereby changing topology. 
Of course we have no obvious control over the
asymptotic particle number of the initial sphaleron-like 
configuration; however, by using suitable stochastic
sampling techniques, we can systematically
lower the particle number while ensuring a change
of topology. This will allow us to explore the
$\epsilon$-$\nu$ plane and map the region of topology
change, the lower boundary of which should tell 
us a great deal about baryon number violation 
in high energy collisions. 

\section{Topological Transitions}
\hspace{1cm}

Let us now put some flesh on the bones of the
introductory discussion. For simplicity we consider 
the standard model with the Weinberg angle set
 to zero. The resulting spontaneously broken 
$SU(2)$ gauge theory possesses all the relevant 
physics while undergoing notable simplification. 
The action for the bosonic sector of this theory is
\begin{equation}
\label{fourAction}
S = \int dx ^4 ~ \left\{- {1 \over 2} {\rm Tr}\,F_{\mu \nu}
    F^{\mu \nu} + D_{\mu} \Phi^\dagger D^{\mu} \Phi  - \lambda
   (\Phi^\dagger \Phi -1 )^2 \right\} \ ,
\end{equation}
where the indices run from $0$ to $3$ and where
\begin{eqnarray}
  F_{\mu\nu} &=& \partial_\mu A_\nu -  \partial_\nu A_\mu
  - i [A_\mu,A_\nu] 
\\
  D_\mu \Phi &=&  (\partial_\mu - i A_\mu) \Phi \ .
\end{eqnarray}
We use the standard metric $\eta={\rm diag}(1,-1,-1,-1)$,
and have eliminated several constants by a suitable choice 
of units. We have also set $g=1$, but  we shall restore the 
gauge coupling to its physical value of $g=0.652$ when 
needed. For our numerical work we take $\lambda=0.1$,
which corresponds to a Higgs mass of about $M_H=
72 \, {\rm GeV}$. 

To yield a computationally manageable system, we work
in the spherical {\it Ansatz} of Ref.~\cite{ry88} in which 
the gauge and Higgs fields are parameterized in terms 
of six real functions $a_0\, ,\,a_1\, ,
\, \alpha\, , \, \beta\, , \, \mu\ {\rm and}\ \nu\ {\rm of}\ r\
{\rm and}\ t$:
\begin{eqnarray}
\label{sphao}
  A_0({\bf x},t) &=& \frac{1}{2 } \, a_0(r,t)
  \lpmb\sigma \cdot {\bf\hat x}
\\
\label{sphai}
  A_i({\bf x},t) &=& \frac{1}{2 } \, \big[a_1(r,t)
  \lpmb\sigma \cdot {\bf\hat x}
  \hat  x^i+\frac{\alpha(r,t)}{r}(\sigma^i- \lpmb\sigma
  \cdot {\bf\hat x}\hat x^i)
  +\frac{1+\beta(r,t)}{r}\epsilon^{ijk}\hat x^j\sigma^k\big]
\\
\label{sphh}
  \Phi({\bf x},t) &=&   [ \mu(r,t) + i \nu(r,t)\lpmb\sigma
  \cdot {\bf\hat x} ] \xi  \ ,
\end{eqnarray}
where ${\bf \hat x}$ is the unit three-vector in the radial direction
and $\xi$ is an arbitrary two-component complex unit vector.
For the \hbox{4-dimensional} fields to be regular at the origin, 
$a_0$, $\alpha$, $a_1 - \alpha/r$, $(1+\beta)/r$ and $\nu$ 
must vanish like some appropriate power of $r$ as $r \to 0$.

These spherical configurations reduce the system to an
effective \hbox{1+1 dimensional} theory whose action  
can be obtained by inserting (\ref{sphao})-(\ref{sphh}) 
into (\ref{fourAction}), from which one obtains\cite{ry88}
\begin{eqnarray}
\label{effAction}
\nonumber
  S =  4\pi \int dt\int^\infty_0dr  &&\bigg[-\frac{1}{4}
  r^2f^{\mu\nu}f_{\mu\nu}+D^\mu \chi^* D_\mu \chi
  + r^2 D^\mu\phi^* D_\mu\phi
\\ 
  && -\frac{1}{2 r^2}\left( ~ |\chi |^2-1\right)^2
  -\frac{1}{2}(|\chi|^2+1)|\phi|^2 -  {\rm Re}(i \chi^* \phi^2)
\\ \nonumber
  && -\lambda  \, r^2 \, \left(|\phi|^2 - 1\right)^2 ~ \bigg] \ ,
\end{eqnarray}
where the indices now run from $0$ to $1$ and 
are raised and lowered with
$\eta_{\mu\nu}={\rm diag}(1,-1)$, and where
\begin{eqnarray}
\label{defConva}   
  f_{\mu\nu}&=& \partial_\mu a_\nu-\partial_\nu a_\mu\
\\
\label{defConvb} 
  \chi &=&\alpha+i \beta
\\
\label{defConvc} 
  \phi &=& \mu+i \nu\
\\
\label{defConvd}  
  D_\mu \chi &=& (\partial_\mu- i   \, a_\mu)\chi
\\
\label{defConve}
  D_\mu \phi&=& (\partial_\mu - \frac{i}{2}  \, a_\mu)\phi\ .
\end{eqnarray}
This is an effective \hbox{2-dimensional} $U(1)$
gauge theory spontaneously broken by two scalar
fields. It possesses the same rich topological structure
as the  full \hbox{4-dimensional} theory and provides 
an excellent testing ground for numerical exploration.

Vacuum states of the effective \hbox{2-dimensional}
theory are characterized by $|\chi |= |\phi|=1$ and
$i\chi^* \, \phi = -1$ (as well as $D_\mu \chi=
D_\mu \phi=0)$. The vacua then take the form
\begin{eqnarray}
\label{amuvac}
 a_{\mu \, \rm vac} &=&  \partial_\mu \Omega
\\
\label{chivac}
  \chi_{\rm vac} \, &=& -i \, e^{i \Omega}
\\
\label{phivac}
  \phi_{\rm vac} \, &=& \pm \, e^{i \Omega/2} \ ,
\end{eqnarray}
where  the gauge function $\Omega=\Omega(r,t)$ 
is required to vanish at $r=0$ to ensure regularity 
of the \hbox{4-dimensional}  fields. Furthermore, 
like the full  \hbox{4-dimensional} theory, these
vacua still possesses nontrivial topological 
structure. Compactification of \hbox{3-space} requires
that $\Omega(r,t) \to 2\pi n$ as $r \to \infty$, in 
which case the winding number of such vacua 
in the the $a_0=0$ gauge is simply the integer 
$n$.  Note that as $r$ varies from zero to infinity, 
$\chi$ winds $n$ times around the unit  circle 
while $\phi$ only winds by half that amount. 

Since the winding number is a topological 
invariant, a continuous path connecting two 
inequivalent vacua must at some point 
leave the manifold of vacuum configurations. 
Along this path there will be a configuration 
of maximal energy, and of all such maximal
energy configurations there exists a unique
one of minimal energy\cite{KM}. This 
configuration is called the sphaleron and 
may conveniently be parameterized by
\begin{eqnarray}
\label{sphSphal}
\nonumber 
  a^\mu_{\rm sph}(r)&=&0  \ ,
\\ 
  \chi_{\rm sph}(r) &=&i [2 f(r)-1] 
\\ \nonumber  
  \phi_{\rm sph}(r)&=&i h(r)   
\end{eqnarray}
where the profile functions $f$ and $h$
vanish at $r=0$, tend to unity as  $r\to \infty$
and are otherwise determined by minimizing
the energy functional. The energy of the
sphaleron depends very weakly on the Higgs
mass and is about $M_w/g^2 \sim 10~{\rm TeV}$,
or $\epsilon=4\pi \, (2.54)$ in the units we are 
using.

\begin{figure}
\centerline{
\epsfxsize=100mm
\epsfbox{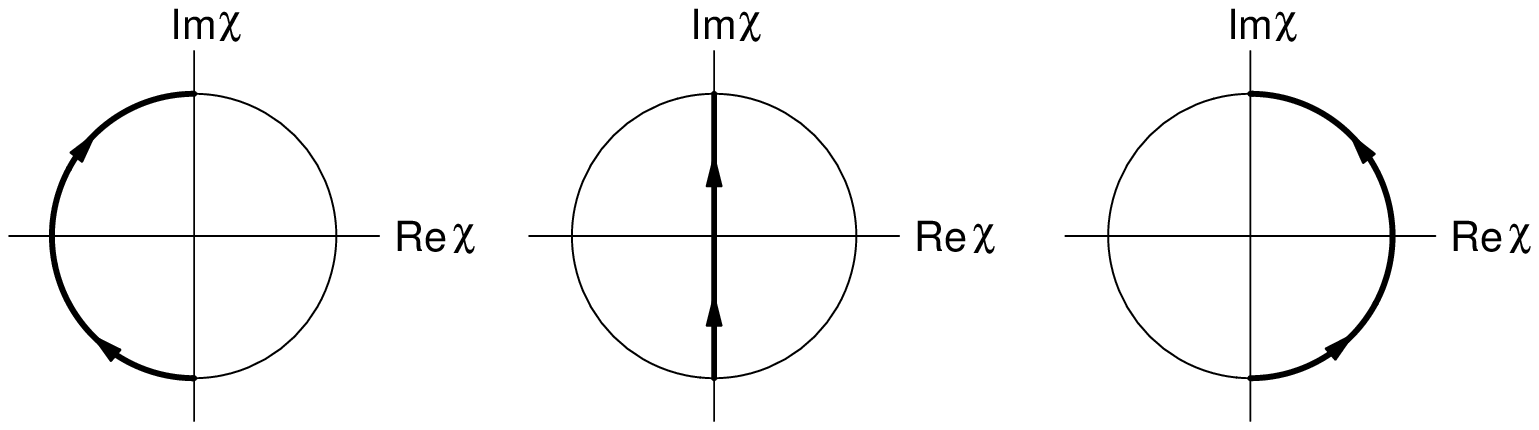}
}
\centerline{
\epsfxsize=105mm
\epsfbox{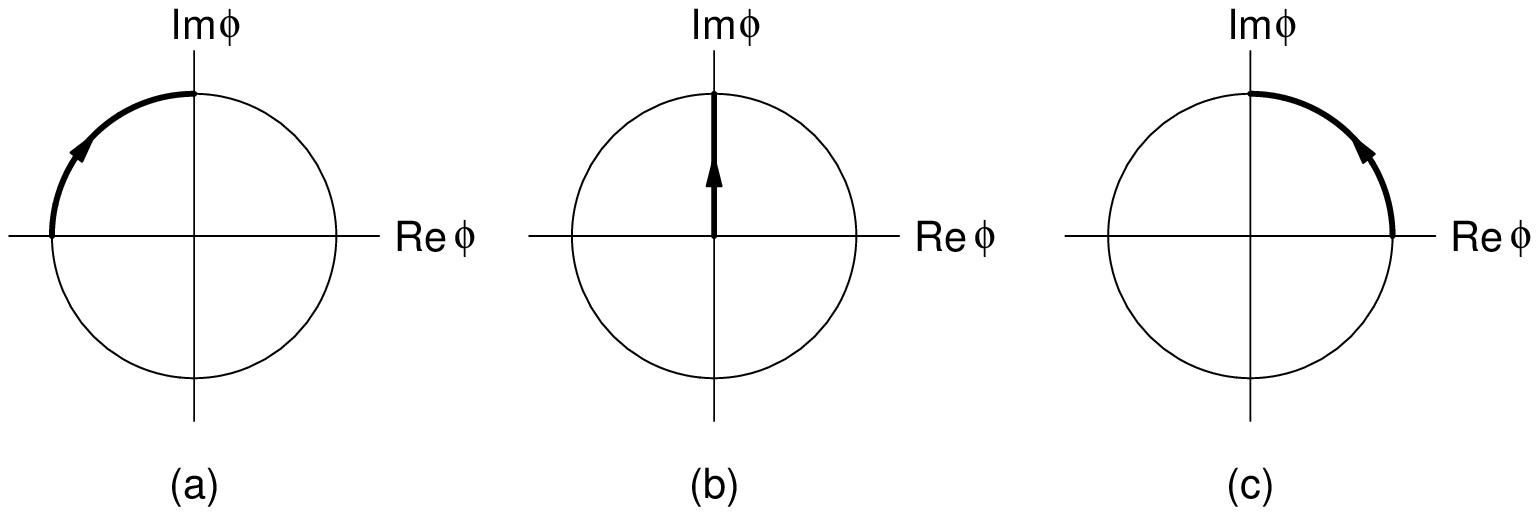}
}
\caption{\tenrm
The $\chi$ and $\phi$ fields for a vacuum-to-vacuum 
topology changing transition in a gauge inconsistent
with compactified \hbox{3-space}.
The scalar fields are traced in the complex plane as 
the the spatial coordinate spans the entire axis.  
Figs.~(a) and (c) represent two inequivalent topological
vacua while (b) is the sphaleron barrier separating them.
}
\end{figure}

While this form of the sphaleron in which $a_\mu$
vanishes and $\chi$ and $\phi$ are pure imaginary
is convenient for numerical work, it does have a 
slight peculiarity that we briefly mention. Recall 
that compactification of \hbox{3-space} requires 
the gauge function $U $ to approach an 
even multiple of $ 2\pi $ as $r \to \infty$. It is
possible to relax this restriction, and it will often
be convenient to choose a gauge in which
$U \to (2n+1) \pi$ as $r \to \infty$, in which 
case $\chi_{\rm vac} \to i$ and $\phi_{\rm vac} 
\to \pm i$. This is precisely the large-$r$ boundary 
condition of the sphaleron, which illustrates 
that (\ref{sphSphal}) is inconsistent with 
spatial compactification. There is of course 
nothing wrong with this, and a topological transition 
of unit winding number change in this gauge
is illustrated in Fig.~1. Rather than $\chi$ winding
once around the unit circle, it instead winds 
over the left hemisphere before the transition
and over the right after the transition.  The total 
phase change is still $2\pi$, as it must be since 
this is a gauge invariant quantity. 

Throughout most of this paper we shall use a gauge 
consistent with (\ref{sphSphal}) in which space
cannot be compactified. From a computational
perspective, this will make perturbations about 
the sphaleron more easily parameterized. There will,
however, be times in which it is more convenient
to impose spatial compactification, but we will
always alert the reader to such a change of gauge.

\section{Classical Evolution}
\hspace{1cm}

So far we have primarily been considering topology
changing sequences of configurations, not 
necessarily solutions of the equations of motion. 
Now we turn to the classical evolution of the
system. We will consider solutions that 
linearize in the distant past and future, and  
hence ones that asymptote to specific topological 
vacua. This allows us to define the incident particle
number, and it makes clear what is meant by 
topology change of a classical solution 
(namely, the change in winding number of 
the asymptotic vacua). 

The field equations are coupled nonlinear
particle differential equations, and  must
be solved computationally on the lattice. But 
before we present our  numerical procedure, 
we first formulate the problem in the continuum. 
The equations of motion resulting from the 
action (\ref{effAction}) are
\begin{equation}
\label{fEq}
  \partial^\mu(r^2f_{\mu\nu})=i \left[D_\nu \chi^*\chi-\chi^*
  D_\nu\chi \right] + \frac{i }{2}\,   \, r^2 \left[D_\nu
  \phi^*\phi-\phi^*D_\nu\phi\right]
\end{equation}
\begin{equation}
\label{chiEq}
  \left[D^2+\frac{1}{r^2}(|\chi|^2-1) + \frac{1}{2}\,  
  |\phi |^2 ~ \right]\chi=-\frac{i}{2}\,  \, \phi^2
\end{equation}
\begin{equation}
\label{phiEq}
  \left[D^\mu r^2 D_\mu+\frac{1}{2}(|\chi|^2+1) +
  2\lambda r^2 \left(|\phi|^2- 1 \right)
  \right] \phi= i \, \chi \phi^* \ .
\end{equation}
The $\nu=0$ equation in (\ref{fEq}) is not dynamical 
but is simply the Gauss's law constraint. 

To solve
these equations, we must supplement them with 
boundary conditions. The conditions at $r=0$ can 
be derived by requiring the \hbox{4-dimensional} 
fields to be regular at the origin. The behavior as 
$r\to 0$ must be
\begin{eqnarray}
\label{zerora}
\label{zeroazero}
  a_0 &=& a_{0,1} r +  a_{0,3} r^3 +\dots 
\\ 
\label{zerorb}
  a_1&=& a_{1,0} + a_{1,2} r^2 + \dots 
\\
  \alpha&=& \alpha_1 r + \alpha_3 r^3  +\dots 
\\
  \beta&=&-1 + \beta_2 r^2 + \dots 
\\
  \mu&=&\mu_0 + \mu_2 r^2 + \dots 
\\ 
\label{zeronu}
  \nu&=&\nu_1 r + \nu_3 r^3 + \dots \ ,
\end{eqnarray}
where the coefficients of the $r$-expansion are 
undetermined functions of time. The $r$-behavior
of the various terms are determined by the
requirement that it has the appropriate power
of \hbox{$r=(x^2+y^2+z^2)^{1/2}$} to render
the \hbox{4-dimensional} fields analytic in terms 
of $x$, $y$ and $z$. For example, $a_0$ must
be odd in $r$ since $A_0$ is proportional to 
$a_0 \lpmb{\sigma} \cdot {\bf \hat x}= (a_0/r) 
\lpmb{\sigma} \cdot {\bf x}$. In terms of $\chi$
and $\phi$ the boundary conditions at $r=0$ 
become 
\begin{eqnarray}
\label{abc}
  a_0(0,t) &=& 0 
\\ 
\label{chibc}
  \chi(0,t)&=&- i  
\\
\label{rephibc}
  {\rm Re}\, \partial_r\phi(0,t)&=&0  
\\
\label{imphibc} 
  {\rm Im} \, \phi(0,t)&=& 0 \ .
\end{eqnarray}
There is another $r=0$ boundary condition which 
arises from the requirement that $a_1-\alpha/r$ 
be regular as $r \to 0$. This condition can be 
written as $a_{1,0}=\alpha_1$, and once imposed 
on initial configurations it remains satisfied 
at subsequent times because of Gauss's law. 

We turn now to the large-$r$ boundary conditions.
Finite energy configurations must approach pure
vacuum at spatial infinity, and we may choose a
gauge in which 
\label{bclarger}
\begin{eqnarray}
\label{alarger}
  a_\mu(r,t) & \to & 0 
\\
\label{chilarger}
  \chi(r,t)  &\to & i  
\\
\label{philarger}
  \phi(r,t) &\to& i  
\end{eqnarray}
as $r\to\infty$. This choice of gauge does not admit
spatial compactification, but nonetheless it is 
numerically conveniently since it is consistent with 
the simple parameterization of the sphaleron (\ref{sphSphal}).  
At times we will choose a gauge consistent with spatial 
compactification in which  $\chi(r,t) \to -i$ and 
$\phi(r,t) \to 1$ as $r \to \infty$, but unless otherwise 
specified we will take the large-$r$ boundary 
conditions to be (\ref{alarger})-(\ref{philarger}). 

The field equations (\ref{fEq})-(\ref{phiEq}), together
with boundary conditions (\ref{abc})-(\ref{philarger}), may now
be used to evolve initial profiles and investigate their subsequent
topology change. The evolution is performed by discretizing
the system using the methods of  lattice gauge theory, in which
we subdivide the $r$-axis into $N$ equal intervals of length
$\Delta r$ with finite extent $L=N \Delta r$ (in our numerical 
simulations we take $N=2239$ and $\Delta r = 0.04$). The
field theoretic system then becomes finite and may be
solved using standard numerical techniques. 

The  fields $\chi(r,t)$ and $\phi(r,t)$ become 
discrete variables $\chi_i(t)$ and $\phi_i(t)$ associated 
with the lattice  sites $r_i= i \Delta r$ where $i=0 \cdots N$.
The continuum boundary conditions render the variables
at the spatial end-points nondynamical, taking the values
\hbox{$\chi_0=-i$},  \hbox{$\chi_N=i$} and \hbox{$\phi_N=i$}
(the value of $\phi_0$ will be discussed momentarily). The 
time component of the gauge field $a_0(r,t)$ is also 
associated  with the lattice sites and is represented 
by the variables $a_{0,i}(t)$ with $i=0 \cdots N$. We
will usually work in the temporal gauge in which $a_{0,i}=0$, 
and we will not concern ourselves with this degree of freedom. 

The spatial components of the gauge field $a_1(r,t)$
become discrete variables associated with the oriented 
links of the lattice, and we represent them by $a_{1,i}(t) 
\equiv a_i(t)$ located at positions $r_{i+1/2}=(i+1/2)\Delta r$
with $i=0\cdots N-1$. The covariant spatial derivatives
become covariant finite difference operators that are
also associated with the links, e.g.
\begin{equation}
\label{Drdisc}
  D_r \phi \to { \exp[- i a_i\, \Delta r/2] \,
  \phi_{i+1}- \phi_i \over \ \Delta r}
  {}~~~~~~~ i=0 \cdots N-1\ .
\end{equation}
where $a_i$ is short-hand notation for  $a_{1,i}$.

It is now straightforward to discretize the action (\ref{effAction})
in a manner that still possesses an exact local gauge invariance. 
But first, we need to state the restriction on $\phi_0(t)$
corresponding to the boundary conditions 
(\ref{rephibc}) and (\ref{imphibc}). Since $a_1$ is real,
we can write these boundary conditions in a 
covariant fashion by requiring the real part of $D_r\phi$
and the imaginary part of $\phi$ to vanish at $r=0$. Using
the discretized operator (\ref{Drdisc}), we can then solve 
this boundary condition for $\phi_0$ to obtain 
\begin{equation}
\label{phi0bc}
  \phi_0 =  {\rm Re} [ \exp(-i \,a_0\, \Delta r/2) \phi_1] \ ,
\end{equation}
where $a_0$ is the value of $a_{1,i}$ at $i=0$ and should
not be confused with the time-like vector field.  This now 
allows us to eliminate $\phi_0$ from the list of
dynamical variables.

Finally, the discretized Lagrangian becomes
\begin{eqnarray}
\label{discL}
\nonumber
  L &=&4\pi \, \sum_{i=0}^{N-1} \bigg\{\frac{r^2_{i+1/2} }{2} \,
  \bigg(\partial_0 a_i   -\frac{a_{0,i+1}-a_{0,i}}{\Delta r}
  \bigg)^2 - \frac{ |\exp(-i\,a_i\, \Delta r)
  \chi_{i+1} -\chi_i|^2}{\Delta r^2} \bigg\} \Delta r
\\ \nonumber && \hskip-0.5in
  + \,4\pi \, \sum_{i=1}^{N-1} \bigg\{ |(\partial_0-i
  a_{0,i})\chi_i|^2+r^2_i \, |(\partial_0-{i a_{0,i} 
  \over 2})\phi_i|^2 - r^2_{i+1/2}  \, \frac{|\exp({-i\,a_i\,
  \Delta r / 2}) \phi_{i+1} -\phi_i|^2}{\Delta r^2 }
\\ &&\hskip-0.5in
  -\frac{1}{2} \, (|\chi_i|^2+1)|\phi_i|^2 -  {\rm Re}(i\chi_i^*
  \phi_i^2) - {1 \over 2 r^2_i}(|\chi_i|^2-1)^2  - \lambda
  r^2_i \, (|\phi_i|^2-1)^2  \bigg\} \Delta r
\\ \nonumber &&\hskip-0.5in
  - \,4\pi \, r^2_{1/2}{[{\rm Im}(\exp(-i\,a_0\, \Delta r / 2)
  \phi_1)]^2 \over  \Delta r} \ ,
\end{eqnarray}
and the system may now be evolved using  standard 
numerical techniques of ordinary differential equations. 
The Lagrangian (\ref{discL}) is actually of a Hamiltonian 
type with no dissipative terms, so it is convenient to 
use the  leapfrog algorithm to perform the numerical 
integration. 

We do not have space to outline this well known 
computational procedure, so instead we simply state 
some of its more attractive features. First, the 
algorithm is second order accurate (i.e. the error 
from time discretization is of order $(\Delta t)^3$ 
in the individual steps and of order $(\Delta t)^2$ 
in an evolution of fixed length). Second, energy is 
exactly conserved in the linear regime, a desirable 
feature when  pulling out the particle number. And 
finally, the algorithm possesses an exact discretized-time 
invariance, which is important since we are interested 
in obtaining the time reversed solutions starting from 
perturbations about the sphaleron. Of course these last 
two properties hold exactly only up to round-off errors, 
which can be made quite small by using double precision 
arithmetic.

\section{The Initial Configuration: Perturbation About
the Sphaleron }
\hspace{1cm}

We are now ready to continue our investigation 
into the connection between the incident particle
number of a classical solution and subsequent
topology change. We could proceed by 
parameterizing linear incoming configurations
of known particle number, but it would be extremely
difficult to arrange the classical trajectory to 
traverse the sphaleron barrier. If we failed to 
see topology change for a given initial 
configuration, we could never be sure whether 
it was simply forbidden in principle by the choice 
of incident particle number, or simply because 
the initial trajectory was pointed the wrong 
direction in field space. 

To alleviate this difficulty, we have chosen 
to evolve initial configurations at or near the 
moment of topology change, and when the 
linear regime is reached the particle number 
will be extracted in the manner explained 
shortly. The physical process of 
interest is then the time reversed solution 
that starts in the linear regime with known 
particle number and subsequently proceeds 
over the sphaleron barrier. Of course we must
explicitly check whether topology change
in fact occurs, but we have found that 
it usually does. 
Fig.~2 illustrates the numerical evolution 
of the $\chi$ field for a typical topology 
changing solution obtained in this manner. 
The modulus of  $\chi$  is represented
by the height of the surface, while the phase
is color coded (but unfortunately we can only 
reproduce the figure in gray scale).  
We have reverted to a gauge in which $\chi_N=-i$ 
and $\phi_N=1$, consistent with spatial compactification,
and in which the incoming state has no winding and
the outgoing state has unit winding number. The
topology change 
is represented by the persistent strip of $2\pi$ 
phase change near the origin after the transition. 

\begin{figure}
\centerline{
\epsfxsize=70mm
\epsfbox{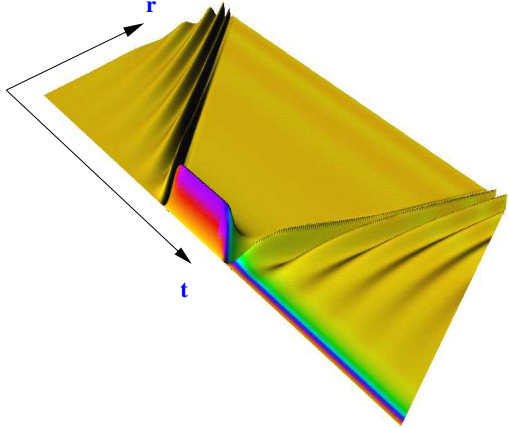}
}
\caption{\tenrm
Topology changing transition: behavior of the $\chi$ field
obtained the time reversal procedure described
in the text. The various shades of gray code the phase of the
complex field. The field starts as an excitation about the
trivial vacuum, passes over the sphaleron and then emerges as an
excitation about the vacuum of unit winding. Note the persistent
strip of $2\pi$ phase change near $r=0$ after the wave bounces
off the origin. 
}
\end{figure}

We turn now to parameterizing initial
configurations. For classical solutions  that 
dissipate in the past and future, topology 
change (and hence baryon number violation) 
is characterized by zeros of the Higgs 
field\cite{fggrs}.
For such topology changing solutions in the
spherical {\it Ansatz}, the $\chi$ field, which 
parameterizes the transverse gauge 
degrees of freedom, must also vanish 
at some point in its evolution. 
However, unless the transition
proceeds directly through the sphaleron, 
the zeros of $\phi$ and $\chi$ need not
occur simultaneously, and for convenience 
we shall choose to parameterize the initial 
configuration  at the time in which $\chi$ 
vanishes for some nonzero $r$. Furthermore,
we can exhaust the remaining gauge 
freedom by taking the initial $\chi$ to be 
pure imaginary. We thus parameterize 
the initial conditions as an expansion in 
terms of some appropriate complete set
with coefficients $c_n$, consistent only with 
the boundary conditions and the requirement
that $\chi$ be pure imaginary with a zero
at some $r>0$. 

We choose to parameterize initial conditions
in terms of perturbations about the sphaleron 
given by linear combinations of spherical Bessel 
functions consistent with the small-$r$ behavior 
(\ref{zeroazero})-(\ref{zeronu}). We only need the first 
three functions
\begin{eqnarray}
\label{jzero}
  j_0(x) &=& \frac{\sin x}{x}   
\\
\label{jone}
  j_1(x) &=& \frac{\sin x}{x^2} - \frac{\cos x}{x}   
\\
\label{jtwo}
  j_2(x) &=&   \left(\frac{3}{x^3} - \frac{1}{x}\right) \sin x -
  \frac{3}{x^2} \cos x  \ ,
\end{eqnarray}
since $j_0(x)\sim 1$, $j_1(x)\sim x$ and $j_2(x)\sim x^2$
at small $x$. We also require the perturbations to vanish
at $r=L$ consistent with the large-$r$ boundary conditions
(\ref{alarger})-(\ref{philarger}). We then parameterize
perturbations about the sphaleron in terms of 
$j_{nm}(r)=j_n(\alpha_{nm}r)$ with $n=0,1,2$, 
where $\alpha_{nm}$ with $m=1,2,\cdots\,$ are the 
zeros of $j_n(x)$. We are thus led to parameterize the
initial conditions as

\begin{eqnarray}
\label{chiparam} 
  \chi(r,0) &=& 
  \chi_{\rm sph}(r) + i \sum_{m=1}^{N_{\rm sph} } \, c_{1 m}\
  j_{2 m}(r) 
\\
\label{phiparam} 
  \phi(r,0) &=& \phi_{\rm sph}(r) + \sum_{m=1}^{N_{\rm sph} } 
  \, c_{2 m}\ j_{0 m}(r)+ i \sum_{m=1}^{N_{\rm sph} } \, 
  c_{3 m}\  j_{1 m}(r)
\\
\label{pichiparam} 
  \dot \chi(r,0) &=& \sum_{m=1}^{N_{\rm sph} } \, c_{4 m}\
   j_{1 m}(r)+ i \sum_{m=1}^{N_{\rm sph} } \, c_{5 m}\  
  j_{2 m}(r)
\\
\label{piphiparam} 
  \dot \phi(r,0) &=&  \sum_{m=1}^{N_{\rm sph} }\, 
  c_{6 m}\ j_{0 m}(r)+ i \sum_{m=1}^{N_{\rm sph} } \, 
  c_{7 m}\  j_{1 m}(r)   
\\
\label{aparam}
  a_1(r,0) &=& \sum_{m=1}^{N_{\rm sph} } \,c_{8 m}\ 
  j_{2 m}(r) \ ,
\end{eqnarray}
where $\chi_{\rm sph}$ and $\phi_{\rm sph}$ are
the sphaleron profiles, and where the sum is cut
off at \hbox{$N_{\rm sph} \le N$}. To avoid exciting short
wave length modes corresponding to lattice
artifacts, we shall take $N_{\rm sph} \sim N/50$ 
(in our numerical work, $N_{\rm sph} =50$ for 
$N=2239$). 

We have used continuum notation, but 
(\ref{chiparam})-(\ref{piphiparam}) is to be thought 
of as defining $\chi$ and $\phi$ on the lattice sites 
$r_i$ and $a_1$ on the links $r_{i+1/2}$. The time 
derivative of $a_1$ is to be determined by Gauss's 
law. 

\section{Normal Modes and  Particle Number}
\hspace{1cm}

We are now in a position to discuss the
manner in which the asymptotic particle
number is to be extracted.
Recall that once the system has reached 
the linear regime it can be represented
as a superposition of normal modes,
and the particle number can be defined 
as the sum of the squares of the 
normal mode amplitudes.
Since we have put the system on a
lattice, we should properly calculate
these amplitudes using the exact 
normal modes of the discrete system. 
However, since our lattice is very
dense ($N=2239$ with $\Delta r=0.04$),
it suffices to project onto the normal 
modes of the corresponding continuum 
system of finite extent $L=N\Delta r$,
the advantage being that we can
solve for the continuum normal modes 
analytically. 
We have checked that this procedure 
agrees extremely well with projecting 
onto normal modes of the discrete system 
(obtained numerically), so for clarity 
we present only the continuum modes. 

It is convenient to work in terms of the gauge 
invariant variables of Ref.~\cite{gi}. We write 
the fields $\chi$ and $\phi$ in polar form,
\begin{eqnarray}
  \chi &=& -i \left[ 1+y \right]\, e^{i\theta} 
\\ \nonumber \\
  \phi &=& \left[ 1+{h \over r} \, \right] \, e^{i\eta} \ ,
\end{eqnarray}
where the variables $y$ and $h$ are gauge invariant.
We can also define the gauge invariant angle
\begin{eqnarray}
\xi=\theta-2\eta \ .
\end{eqnarray}
Finally, in \hbox{1+1} dimensions we can write
\begin{eqnarray}
  r^2 f_{\mu\nu} = - 2 \epsilon_{\mu\nu} \psi
\end{eqnarray}
where $\epsilon_{01}=+1$ and $\mu$, $\nu$ run 
over $0$ and $1$, and where the variable $\psi$ 
is gauge invariant. Rather than working with the 
six gauge-variant degrees of freedom $\chi$, 
$\phi$ and $a_\mu$ we use the four gauge 
invariant variables $\rho$, $\sigma$, $\psi$ 
and $\xi$. 

We wish to find the equations of motion for 
small linearized fluctuations about the vacuum. 
In gauge invariant coordinates the vacuum takes 
the form $h_{\rm vac}=y_{\rm vac}=\psi_{\rm vac}=
\xi_{\rm vac}=0$, and we thus need  only work
to linear order in the variables. From 
Ref.~\cite{gi} the normal mode equations are
\begin{eqnarray}
\label{giSigmaLin} 
 \Biggl(\partial_\mu\partial^\mu + 4\lambda \Biggr)\,h  &=&0
\\
\label{giRhoLin} 
 \Biggl(\partial_\mu\partial^\mu + \frac{1}{2}  +
 \frac{2}{r^2}\Biggr)\,y &=& 0
\\
\label{giPsiLin} 
 \partial^\mu\left\{ \frac{\partial_\mu \psi -  \epsilon_{\mu\nu}
 \partial^\nu \xi}{1 + \frac{1}{4}r^2}\right\} + \frac{2}{r^2} \,
 \psi &=& 0
\\
\label{giXiLin}
 \partial^\mu\left\{ \frac{ \frac{1}{4} r^2 \partial_\mu \xi +
 \epsilon_{\mu\nu}\partial^\nu \psi }{1 + \frac{1}{4}r^2 }
 \right\} + \frac{1}{2} \xi &=& 0 \ .
\end{eqnarray}
Equation (\ref{giSigmaLin}) corresponds to a 
pure Higgs excitation characterized by mass 
$M_H=2\sqrt{\lambda}$, while 
(\ref{giRhoLin})-(\ref{giXiLin}) correspond to
three gauge modes of mass  
$M_W = 1/\sqrt{2}$.\footnote[1]{Upon restoring 
the factors of $g$ and the Higgs vacuum
expectation value $v$, these masses take the standard
form $M_H=\sqrt{2\lambda}\, v$ and $M_W = 
(1/2)g\, v$.}

Note that there are four types of normal modes. 
The first two are easily obtained by solving the 
independent equations  (\ref{giSigmaLin}) and 
(\ref{giRhoLin}), while the last two can be 
found by solving the coupled equations 
(\ref{giPsiLin}) and (\ref{giXiLin}) involving 
$\psi$ and $\xi$. 
A solution in the linear regime can then
be expanded as a combination of these
four modes and the amplitudes $a_{k n}$, 
with $k=1,2,3,4$ specifying the mode
type, extracted. The Higgs and
gauge particle numbers are defined by 
\begin{eqnarray}
  \nu_{\rm higgs} &=& \sum_{n=1}^{N_{\rm mode}}
  |a_{1n}|^2  
\\
  \nu_{\rm gauge} &=& \sum_{n=1}^{N_{\rm mode}}
  \left\{ |a_{2n}|^2 +
  |a_{3n}|^2 +|a_{4n}|^2 \right\} \ ,
\end{eqnarray}
with total particle number given by
\begin{eqnarray}
  \nu = \nu_{\rm higgs} + \nu_{\rm gauge} \ .
\end{eqnarray}
To avoid counting lattice artifacts we take 
the ultraviolet cutoff on the mode sums to 
be given by $N_{\rm mode} \sim N/5$ to 
$N/10$. 

Space does not permit a detailed exposition
of this procedure, and one should consult 
Ref.~\cite{rs} for full details. Here we must be 
content with Fig.~3, which displays the behavior 
of the particle number in the four modes of 
oscillation as a function of time. The initial 
state was a typical perturbation about the 
sphaleron as described in the previous section,
and as it evolves it quickly linearizes and
settles down into a definite asymptotic particle 
number.

\begin{figure}
\centerline{
\epsfxsize=100mm
\epsfbox{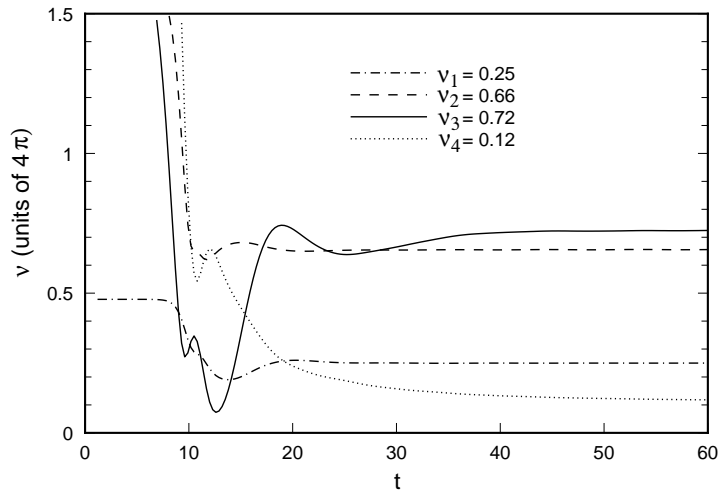}
}
\caption{\tenrm
Decay of a small perturbation about the sphaleron:
behavior of the particle number in the four modes  
as function of time for lattice parameters $N=2239$, 
$\Delta r = 0.04$ and $N_{\rm mode}=200$ with
$\lambda=0.1$. The physical particle numbers are 
obtained by multiplying the asymptotic values in the 
graph by $4\pi/g^2 \sim 30$, which gives 
$N_{\rm higgs}\sim8$ and $N_{\rm gauge}\sim45$, 
for a total physical particle number of $N_{\rm phys}\sim53$.
}
\end{figure}

\section{Stochastic Sampling of Initial  Configurations}
\hspace{1cm}

Recall that our computational strategy consists in 
evolving a configuration near the top of the sphaleron 
barrier until it linearizes, at which point the particle 
number can be extracted and the time reversed 
solution used to generate the topology changing 
process of interest. We can regard the energy 
$\epsilon$ and the asymptotic particle number $\nu$ 
as functions of the parameters $c_n$ that
specify the initial configuration, and by varying
these coefficients we would like to explore 
the  $\epsilon$-$\nu$ plane and attempt to 
map the region of topology change. In
particular, for a given energy $\epsilon$, 
we would like to find the minimum allowed 
particle number $\nu_{\rm min}(\epsilon)$
consistent with a change of topology. If 
this number can be made arbitrarily small,
this would be a strong indication that baryon
number violation would be observable
in a two-particle collision. 

By randomly exploring the initial configuration 
space parameterized by the coefficients $c_n$, 
we would stand little chance of making headway. 
Instead, we shall employ stochastic sampling 
techniques, which are ideal for tackling
this type of multi-dimensional minimization.
Our procedure will be to generate initial
configurations  weighted by  $W = \exp(-F)$ 
with $F=\beta \, \epsilon + \mu \, \nu$,
and by adjusting the parameters $\beta$
and $\mu$ we can explore selected
regions in the $\epsilon$-$\nu$ plane.
In particular, by increasing $\mu$ we 
can drive the system to lower and lower 
values of $\nu$ for a given $\epsilon$.
In our numerical work we typically take
$\beta$ between 50 and 1000 while
$\nu$ ranges between 1000 to 20000.

To generate the desired distributions
we have used a Metropolis Monte-Carlo
algorithm. Starting from a definite 
configuration parameterized by 
$c_n$, we perform an upgrade to 
\hbox{$c_n \to c'_n = c_n +\Delta c_n$} 
where $\Delta c_n$ is Gaussian 
distributed with a mean of about 0.0008. 
We evolve the updated configuration
until it linearizes and then calculate
$\Delta F=\beta \, \Delta \epsilon 
+ \mu \, \Delta \nu$. If the topology
of the physically relevant time reversed
solution does not change, then we
discard the updated configuration.
Otherwise we accept it with conditional 
probability $p={\rm Min}[1,\exp(-\Delta F)]$, 
which is equivalent to always accepting 
configurations that decrease $F$ while 
accepting those that increase $F$  with 
conditional probability  $\exp(-\Delta F)$.

\begin{figure}
\centerline{
\epsfxsize=100mm
\epsfbox{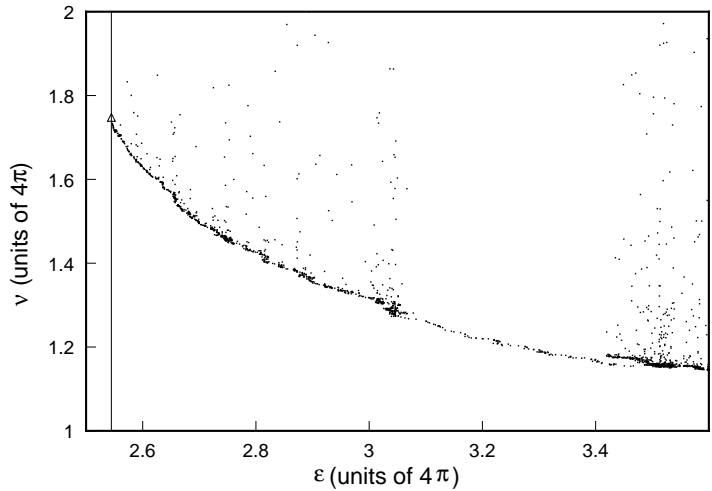}
}
\caption{\tenrm
Monte Carlo results with lattice parameters of $N=2239$,
$\Delta r = 0.04$ (giving L=89.56), $N_{\rm mode}=200$
and $N_{\rm sph}=50$, and with a Higgs self-coupling of
$\lambda=0.1$. The solid line marks the sphaleron
energy $\epsilon_{\rm sph}= 4\pi (2.5426)$, below which
no topology changing process can lie. The triangle represents
the configuration from which we seeded our Monte Carlo
search. To obtain quantities in physical units, multiply the
numbers along the axes by $4\pi/g^2 \sim 30$. The energy axis
extends from about $10 {~\rm TeV}$ to $15 {~\rm TeV}$, while the
particle number axis ranges from about $30$ particles to $60$.
}
\end{figure}

We are now in a position to present our numerical
results. Fig.~4 represents 300 CPU hours 
and involves
30000 solutions (of which only 3000 are shown) 
obtained on the CM-5, a 64 node parallel 
supercomputer.  We have chosen the lattice
parameters \hbox{$N=2239$}, \hbox{$\Delta r=0.04$}, 
with ultraviolet cutoffs determined by 
\hbox{$N_{\rm sph}=50$} and 
\hbox{$N_{\rm mode}=200$}.
The Higgs self-coupling was taken to be
$\lambda=0.1$, which corresponds to a
Higgs mass of $M_H=72 \, {\rm GeV}$.

We have managed to produce a marked decrease
of about 40\% in the minimum particle number 
$\nu_{\rm min}(\epsilon)$, which is approximated
by the lower boundary in the Fig.~4. Nowhere,
however, in the explored energy range does $\nu$
drop below $4\pi$, or in physical units the
incident particle number $N \ge 30$ for energy 
$E \le 15 \, {\rm TeV}$ (the outgoing particle
number tends to be about 50 to 100). This is 
a far cry from two incoming particles which would 
be necessary to argue that baryon number becomes 
unsuppressed in high energy collisions. 

The complex nature of the solution space can
be illustrated by the break in population
density between $\epsilon/4\pi \sim 3 $
and $\epsilon/4\pi \sim 3.4$. In our first 
extended search we did not check whether 
topology change actually occurred, trusting
that the time reversed solutions would continue over
the sphaleron barrier. However, we later found 
an entire region between  $\epsilon/4\pi \sim 3 $
and $\epsilon/4\pi \sim 3.4$ in which the
solutions never left the original topological
sector. We excluded these points and
restarted our search procedure near
$\epsilon/4\pi \sim 3 $. A small discontinuity
in the lower boundary with slightly lower particle 
number was produced, but we have still
managed to approximate $\nu_{\rm min}(\epsilon)$
remarkably well. 

We can extract more information from the 
system by investigating the asymptotic
spectral distribution  $|a_{kn}|^2$ as 
a function of mode number $n$. Before 
we started the search, our seed  configuration
(represented by the triangle in Fig.~4)
linearized into a distribution that was 
heavily peaked about a small mode
number $n_{\rm pk} \sim 50$ (with $\Delta n 
\sim 50 $), corresponding to a frequency of
$\omega_{\rm pk} \sim \pi n_{\rm pk}/L \sim 0.1$. 
After the search the solutions underwent 
a dramatic mode redistribution. The amplitudes 
$|a_{kn}|^2$ of the linear regime peaked at 
higher mode number, $n_{\rm pk} \sim 75-100$,
with a much broader distribution ($\Delta n 
\sim 200$). Clearly our search procedure is 
very efficient in redistributing the mode 
population density.

While $\nu$ remains large throughout 
the energy range we have explored, it 
is interesting to note that $\nu_{\rm min}
(\epsilon)$ maintains a slow but steady 
decrease with no sign of leveling off. 
To obtain an indication of the possible
behavior of $\nu_{\rm min}(\epsilon)$
at higher energies, we performed fits
to our data using functional forms which
incorporate expected analytical properties 
of the boundary of the domain of topology
changing solutions. 
The fits gave a particle number $N=2$ at
energies in the range of $100 \, {\rm TeV}$  
to $450 \, {\rm TeV}$. Of course we must 
explore higher energies before drawing 
define conclusions, but this is at least 
suggestive that particle number might 
at some point become small. 

While the energy range we have explored 
is of limited extent and more numerical
work is clearly called for, it is still
remarkable that we have extracted such 
a wealth of information from such an 
analytically intractable field theoretic 
system. 
Computational
techniques offer considerable promise 
in probing the nonlinear dynamics of 
the standard model, and we fully expect
them to play a prominent role in our
future understanding of high energy
baryon number violation,
both in extending the energy range of
the classically allowed processes and
obtaining information on the classically
forbidden processes below the barrier.

{\bf Acknowledgments}

This research was supported in part under DOE grant
DE-FG02-91ER40676 and NSF grant ASC-940031. We wish to thank
V.~Rubakov for very interesting conversations which stimulated
the investigation described here, A.~Cohen, K.~Rajagopal and
P.~Tinyakov for valuable discussions, and T.~Vaughan for
participating in an early stage of this work.

\end{document}